 \makeatletter
 \@ifundefined{@parse@version@dash}{%
 \def\@parse@version#1{\@parse@version@0#1}
 \def\@parse@version@#1/#2/#3#4#5\@nil{%
 \@parse@version@dash#1-#2-#3#4\@nil}
 \def\@parse@version@dash#1-#2-#3#4#5\@nil{%
   \if\relax#2\relax\else#1\fi#2#3#4 }
 }{}
 \makeatother

\documentclass[aps,apl,reprint,groupedaddress]{revtex4-1}

\usepackage{color}
\usepackage[space]{grffile}
\usepackage{float}
\usepackage{lineno}
\usepackage{amstext}
\usepackage{amsfonts}  
\usepackage{amsmath}
\usepackage{siunitx}

\usepackage[version=3]{mhchem}
\usepackage{amssymb}
\usepackage{amsthm}
\usepackage{textgreek}
\usepackage{textcomp}
\usepackage{mathrsfs}
\usepackage{babel}
\usepackage{graphicx}
\usepackage{epstopdf}
\usepackage{color}

\DeclareSIUnit\bar{bar}
\DeclareSIUnit\torr{torr}
\DeclareSIUnit\sccm{sccm}

\renewcommand{\cite}{\citep}

\begin{document}


\title[Mean inner potential of alloyed and strained materials]{Modelling the mean inner potential of alloyed and strained materials}


\author{Marco Schowalter$^1$$^2$}
\email{schowalter@uni-bremen.de}
\author{P. Kruse$^3$}
\author{Andreas Rosenauer$^1$$^2$}

\affiliation{$^1$Institut f\"ur Festk\"orperphysik, Universit\"at Bremen, Otto-Hahn-Allee 1, 28359 Bremen, Germany}
\affiliation{$^2$MAPEX Center for Materials and Processes, Universit\"at Bremen, Bibliothekstra{\ss}e 1, 28359 Bremen, Germany}
\affiliation{$^3$ Robert Bosch GmbH, Wernerstra{\ss}e 51, 70469 Stuttgart, Germany}


\date{\today}

\begin{abstract}
In this publication, we study the influence of strain and alloying on the mean inner potential (MIP) using density functional theory (DFT) within an augmented plane waves plus local orbitals (APW+LO) basis set. Two major effects have been identified allowing to model the influence of strain and alloying on the mean inner potential with a reasonable accuracy. First, alloying for constant volume results in a linear relationship between the MIP and the concentration. Second, the MIP scales with changes in volume as we already pointed out in an earlier publication [{\it M. Schowalter, D. Lamoen, A. Rosenauer, P. Kruse, and
D. Gerthsen, Appl. Phys. Lett. 85, 4938–4940 (2004)}]. Specifically, a linear relationship between MIP and concentration $x$ was found for Al$_x$Ga$_{1-x}$As (nearly no change in lattice parameter), whereas In$_x$Ga$_{1-x}$P and Ge$_{x}$Si$_{1-x}$ (volume changes with concentration $x$) exhibits a clear bowing. The bowing can be modeled by taking the rescaling of the MIP with the varying volume additionally into account. The rescaling could be also used to model the dependence of the MIP on strained binary cells and the density dependence of e.g. amorphous materials. 
\end{abstract}


\maketitle



The mean inner potential (MIP) is defined as the average Coulomb potential 
\begin{equation}
V_0=\frac{1}{\Omega} \int\limits_\Omega V_\text{c}(\vec{r}) d^3r\quad\text{,}
\label{eq:MIP_def}
\end{equation}
where $\Omega$ is the integration volume and $V_\text{c}(\vec{r})$ the Coulomb potential at position $\vec{r}$. The zero point of the Coulomb potential $V_\text{c}$ is chosen such that it is zero at infinite distance from the crystal. Thus, it depends on the crystal structure and the surface of the crystal \cite{SALDIN}. Frequently, it is approximately derived from scattering factors computed for isolated atoms as
\begin{equation}
V_0^\text{iso}=\frac{h^2}{2\pi m e \Omega} \sum\limits_{i} f^\text{el}_i(0) \quad\text{,}
\label{eq:MIPiso}
\end{equation}
with $f^\text{el}_i(0)$ the electron scattering amplitude (taken from e.g. \cite{Doyle1968}) at zero $k$, $i$ running over all atoms within the unit cell as well as $h$, $m$ and $e$ being Planck's constant, the mass and charge of the electron, respectively. In this definition, the effects of structure and surface of the crystal as well as charge redistribution due to bonds on the MIP are not accounted for. Kim et al.\cite{Kim1998} used density functional theory in order to overcome these limitations for MgO, Ge and Si. For the latter, they found that the type of surface or surface relaxation changed the MIP by about 0.2 V, whereas the redistribution of charge resulted in a change by 1.7 V (nearly one order of magnitude larger). Similar results were found for the MIPs of II-VI \cite{SchowalterIIVI} and III-V semiconductors \cite{KruseMIPDFT,SchowalterWurtz} as well as amorphous carbon for various mass densities \cite{SchowalterAmorphC}. Typical precisions for the measurement of MIPs with current techniques (electron holography) are in the range of about 0.1-0.4 V \cite{GAJDARDZISKAJOSIFOVSKA1993285,KRUSE200311,AUSLENDER201918,AUSLENDER2021288}, so that different surfaces are hardly experimentally distinguishable, but the influence of redistribution of charge is well measurable. 

Rosenauer et al. \cite{RosenauerMASA} introduced the concept of modified atomic scattering amplitudes (MASAs) in order to account for the redistribution of charge on the scattering of the chemically sensitive (002) beam in alloyed sphalerite type semiconductors. The MASAs were derived from density functional theory (DFT) calculations using an augmented plane wave plus local orbitals (APW+LO) basis sets. Therein, space is subdivided into muffin-tin spheres around the atom positions with atomic-like orbitals and the interstitial region with plane wave basis sets. For each region contributions to the structure factor were computed and by assigning the fractions of the interstitial part to the individual atomic contributions atomic scattering amplitudes were derived, which were slightly modified, but which accounted for the redistribution of charge. It was found that the structure factor derived from MASAs very well described the structure factor of a full computation. In fact, the MASA concept in principle could be applied for the scattering amplitudes in forward direction and thus, for the computation of the MIP in an analogous manner.

Recently, the influence of alloying and strain on MIP has been adressed by different authors \cite{Mergner_alloying,Denaix_strain}. For alloying, a linear relation between the unalloyed materials was assumed for the interpretation of the electron holography results on p-AlGaAs/n-GaInP solar cell junctions \cite{Mergner_alloying}. Strain dependences of the MIPs for AlN and GaN have been computed by Denaix et al. \cite{Denaix_strain} in order to estimate their found discrepancy between their measurements of the MIP difference between AlN and GaN and their own computed values. 

These works inspired us to investigate the influence of alloying and strain on the MIP and suggest a model, with which both can be taken into account from just the knowledge of the MIP of unalloyed and unstrained materials. The key idea is as follows: 
\begin{enumerate}
\item Assuming that we would have computed MASAs for forward scattering, expanding eq. (\ref{eq:MIPiso}) directly shows that the MIP of an alloyed material $V_0^{A_xB_{1-x}}$ is a linear combination of the MIPs of the unalloyed materials $V_0^A$ and $V_0^B$
\begin{equation}
V_0^{A_xB_{1-x}}=x V_0^A + (1-x) V_0^B\quad{.}
\label{eq:MIPlinear}
\end{equation}
\item Assuming that we would have computed MASAs for forward scattering, equation (\ref{eq:MIPiso}) suggests that the influence of any change in volume on the MIP could be calculated just by dividing by the respective volume. In other words, if the MIP $V_0(\Omega)$ at a volume $\Omega$ of a system is known than the MIP $V_0^\prime(\Omega^\prime)$ of the system at a volume $\Omega^\prime$ can be derived by 
\begin{equation}
V_0^\prime(\Omega^\prime)=\frac{\Omega}{\Omega^\prime}V_0(\Omega)\quad{.}
\label{eq:MIPrescaling}
\end{equation}
\end{enumerate}

This simple approach of course only holds if the influence of any charge redistribution due to alloying, straining, surface relaxation etc. is significantly smaller than the difference to isolated atoms.


In this publication, we computed the MIP for differently strained GaP cells, the composition dependence of the MIP of Al$_x$Ga$_{1-x}$As (nearly no change in volume) and model both by idea (2) and (1), respectively. For the composition dependence of the MIP of In$_x$Ga$_{1-x}$P and Ge$_x$Si$_{1-x}$  both, ideas (1) and (2) need to be combined, as lattice parameter vary with concentration.


We used density function theory within an APW+LO basis set as implemented in the WIEN2k code \cite{WIEN2k}. For all calculations standard settings as automatically suggested by the ''init'' function of WIEN2k were applied. This implies the choice of the generalized gradient approximation for the exchange and correlation part of the potential as parametrized by \cite{PBE} and about 100 k-points or 500 k-points (for 2$\times$1$\times$2 or 1$\times$1$\times$1 supercells, respectively) in the full Brillouin zone. For all computations the tetrahedron integration method was used except for Ge$_x$Si$_{1-x}$, for which a smearing with width 0.005 Ry was necessary.

For the simulation of the impact of strain on the MIP of GaP we chose the same slab geometry as in \cite{KruseMIPDFT,SchowalterIIVI}. In these publications, crystal directions were chosen such that [1-10], [110] and [001] were parallel to the $x$, $y$ and $z$, respectively and an approximately 1~nm large vacuum area was symmetrically added in y-direction. Thus, slabs consisted of 22 atoms with 2 atoms in each monolayer, i.e. one Ga and one P atom. The MIP was then computed from the innermost monolayer as described in \cite{KruseMIPDFT}. 

Alloying was simulated for rather ionically bound In$_x$Ga$_{1-x}$P and the Al$_x$Ga$_{1-x}$As as well as covalently bound Ge$_x$Si$_{1-x}$. For that, 2$\times$1$\times$2 supercells were generated from the simple slabs containing now 88 atoms per slab and 8 atoms per monolayer. As the MIP was computed from the innermost monolayer we decided to fix the target concentration in each monolayer. Thus, concentrations $x$ of 0, 0.25, 0.50, 0.75 and 1.00  could be realized for the In$_x$Ga$_{1-x}$P and the Al$_x$Ga$_{1-x}$As alloy as well as 0, 0.125, 0.25, 0.375, 0.50, 0.625 0.75, 0.875 and 1.00 for Ge$_x$Si$_{1-x}$. Still, the number of possible configurations $N={{n}\choose{k}}^{n_\text{ML}}$, with $n=4$ ($n=8$) and $k \in [0;4]$ ($k \in [0;8]$) and $n_\text{ML}$=11 were too demanding for our computational resources. So we restricted, the realized configurations to $n_\text{ML}$=3 ($n_\text{ML}$=1) innermost layers for InGaP and AlGaAs (GeSi). Outer monolayers were not systematically varied, but for each configuration in the innermost layer a random configuration (with fixed concentration per monolayer) of the outer monolayers was realized. Chosen cells can be found in ref. \cite{Zenodo}. Thus, the MIP of 1, 64, 216, 64 and 1 (1, 8, 28, 56, 70, 56, 28, 8, 1) configurations were computed for InGaP and AlGaAs (GeSi), respectively. For each concentration the MIPs were averaged with Boltzmann distributed weights \cite{SchowalterInGaOstrain}.


\begin{figure}[h!]
    \centering
    \includegraphics[width=0.95\linewidth]{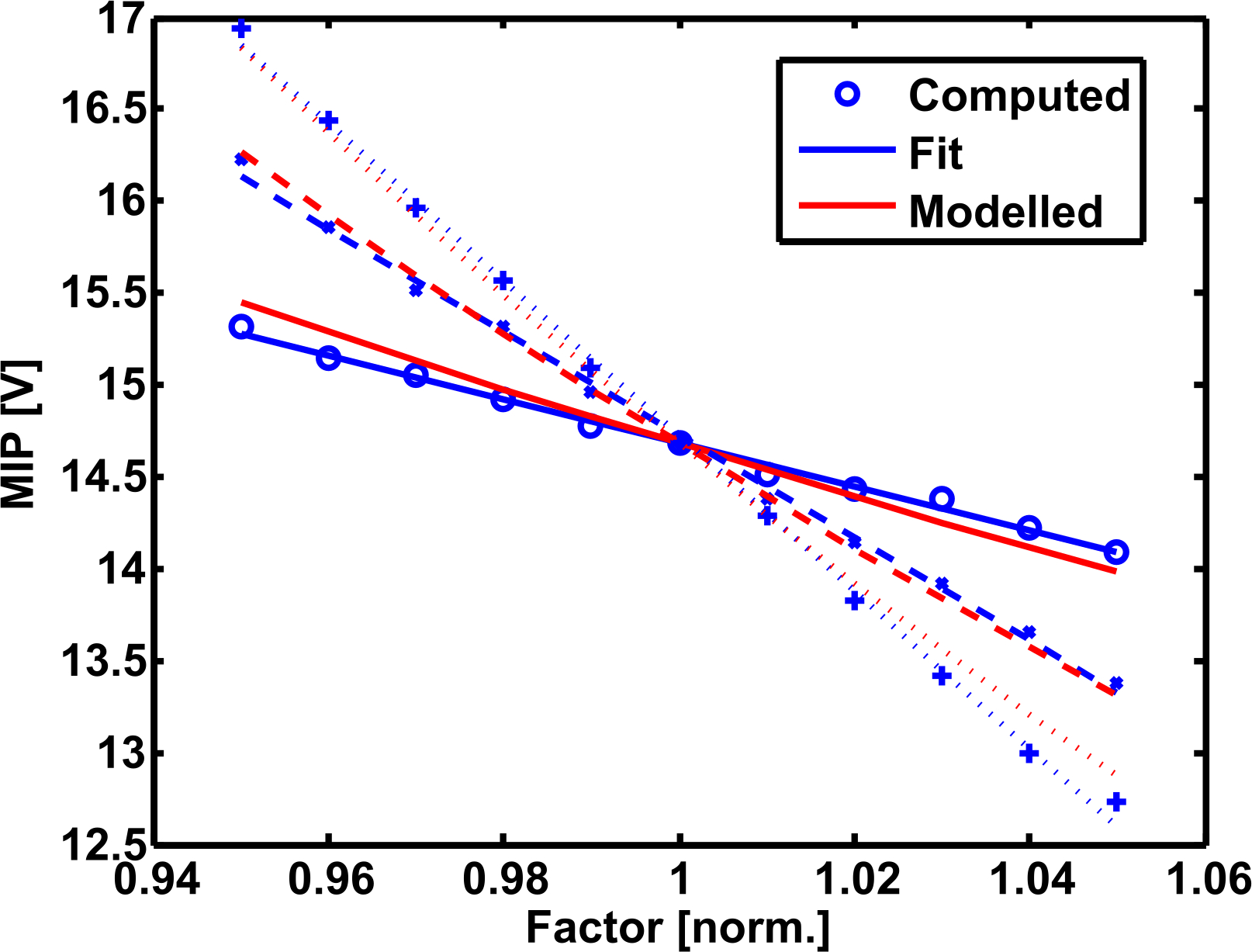}
    \caption{The value of the MIP as function of the factor $f$ with which the lattice parameters of GaP were multiplied. Three different situations were considered: uniaxial strain (only z-direction varied, circles and straight lines), biaxial strain (x and z direction varied, crosses and dashed lines) and hydrostatic strain (x,y,z directions varied, +-signs and dotted lines). Linear relation are found, with different slopes, which can be models reasonably by rescaling the volume (red curves).}
    \label{fig:MIP_strain_GaP}
\end{figure}

 \begin{figure}[h!]
    \centering
    \includegraphics[width=0.9\linewidth]{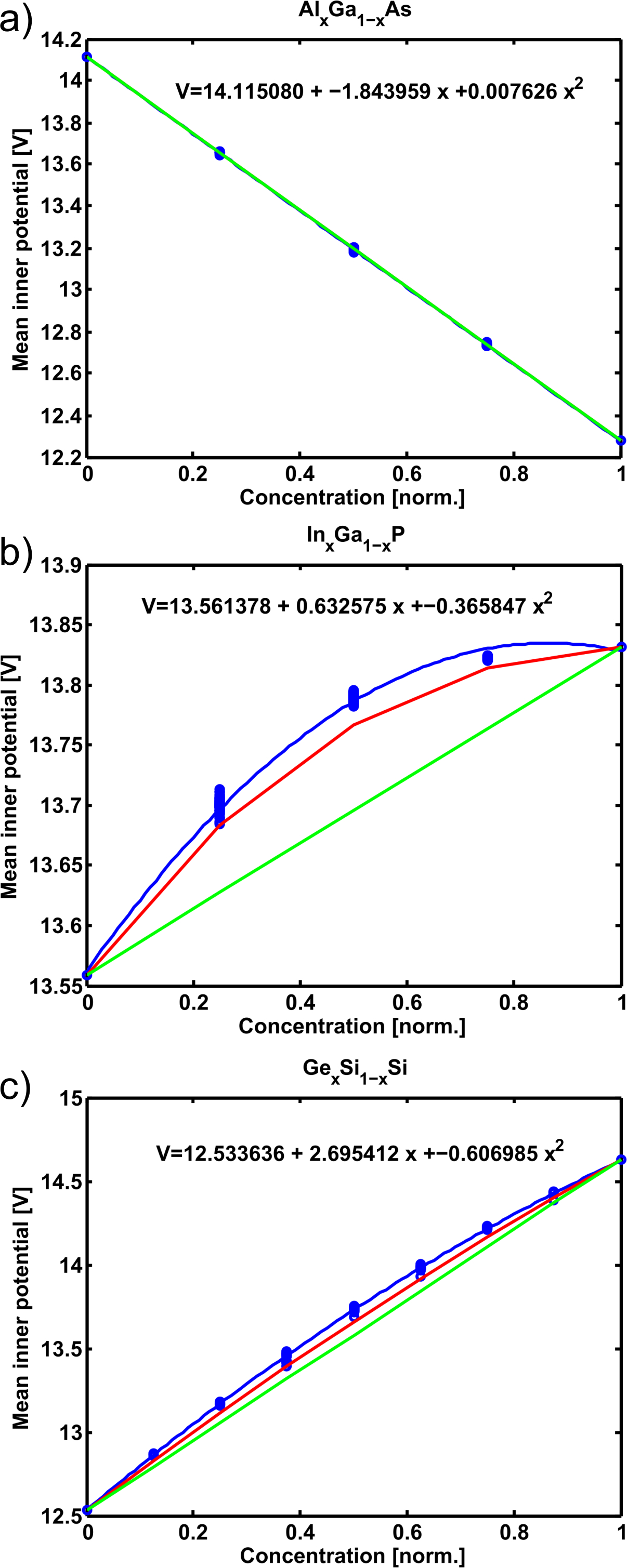}
    \caption{Computed MIPs for a) Al$_x$Ga$_{1-x}$As, b) In$_x$Ga$1-x$P and c) Ge$_x$Si$_{1-x}$ plotted as blue circles. Solid blue lines are second order polynomial fits to the Boltzmann averaged MIPs. The green solid and red dashed line indicate the dependence assuming only linear interpolation and additionally volume rescaling.}
    \label{fig:MIP_alloy_relaxed}
\end{figure}

In a first step, the dependence of the MIP on strain was investigated for GaP using 1$\times$1$\times$1 cells as outlined before. 
For that, the size of the crystalline slab was multiplied by a factor $r$ in $z$ only, in two direction $x$ and $z$ or all three directions $x$,$y$,$z$ in order to mimic uniaxial strain, biaxial strain and hydrostatic strain,
respectively. Fig. \ref{fig:MIP_strain_GaP} shows the dependence of the computed MIP on the multiplicative factor $r$ for uniaxial strain (circles), biaxial (crosses) and hydrostatic strain (''+''). The computed data could be well fitted with linear functions depicted as the respective blue lines. Red lines show the plot of the suggested model as outlined in eq. (\ref{eq:MIPrescaling}). Though the agreement is not perfect, it describes the slope of the linear function under the different strain situations not too bad. Maximum deviations between the proposed model and the fit are 1.1\% (0.17 V), 0.8\% (0.14 V) and 2.2\% (0.28 V) for uniaxial, biaxial and hydrostatic strain, respectively. 
Such deviations can hardly be detected with the precision of current methods such as e.g. electron holography \cite{GAJDARDZISKAJOSIFOVSKA1993285,KRUSE200311,AUSLENDER201918,AUSLENDER2021288}.

Figure \ref{fig:MIP_alloy_relaxed} shows the computed MIP as a function of concentration for unstrained a) Al$_x$Ga$_{1-x}$As, b) In$_x$Ga$_{1-x}$P and c) Ge$_x$Si$_{1-x}$ as blue circles as well as respective fits with a 2nd order polynomial (blue solid lines). For these computations only the change of lattice parameter according to Vegard's law and no strain was applied to the cells. Green and red solid lines indicate the results of the proposed model only considering alloying by eq. (\ref{eq:MIPlinear}) and both equations ((\ref{eq:MIPlinear}) and (\ref{eq:MIPrescaling})), respectively. For  Al$_x$Ga$_{1-x}$As both models agree well with the computations as the lattice parameter of AlAs and GaAs are nearly identical (see table \ref{tab:LattPara}) and thus, the prefactor in eq. (\ref{eq:MIPrescaling}) $\frac{\Omega}{\Omega^\prime}\approx 1$. 
For the remaining In$_x$Ga$_{1-x}$P and Ge$_x$Si$_{1-x}$ systems the lattice mismatch is significantly larger (7.7\%  and 4.2\% compared to 0.1\%) and thus, the rescaling by the prefactor in eq. (\ref{eq:MIPrescaling}) cannot be neglected. The prefactor $r$ leads to a bowing of the MIP with concentration. Again the model with both equations (\ref{eq:MIPlinear}) and (\ref{eq:MIPrescaling}) describes the computed behaviour not too bad. 

\begin{figure}[h!]
    \centering
    \includegraphics[width=0.9\linewidth]{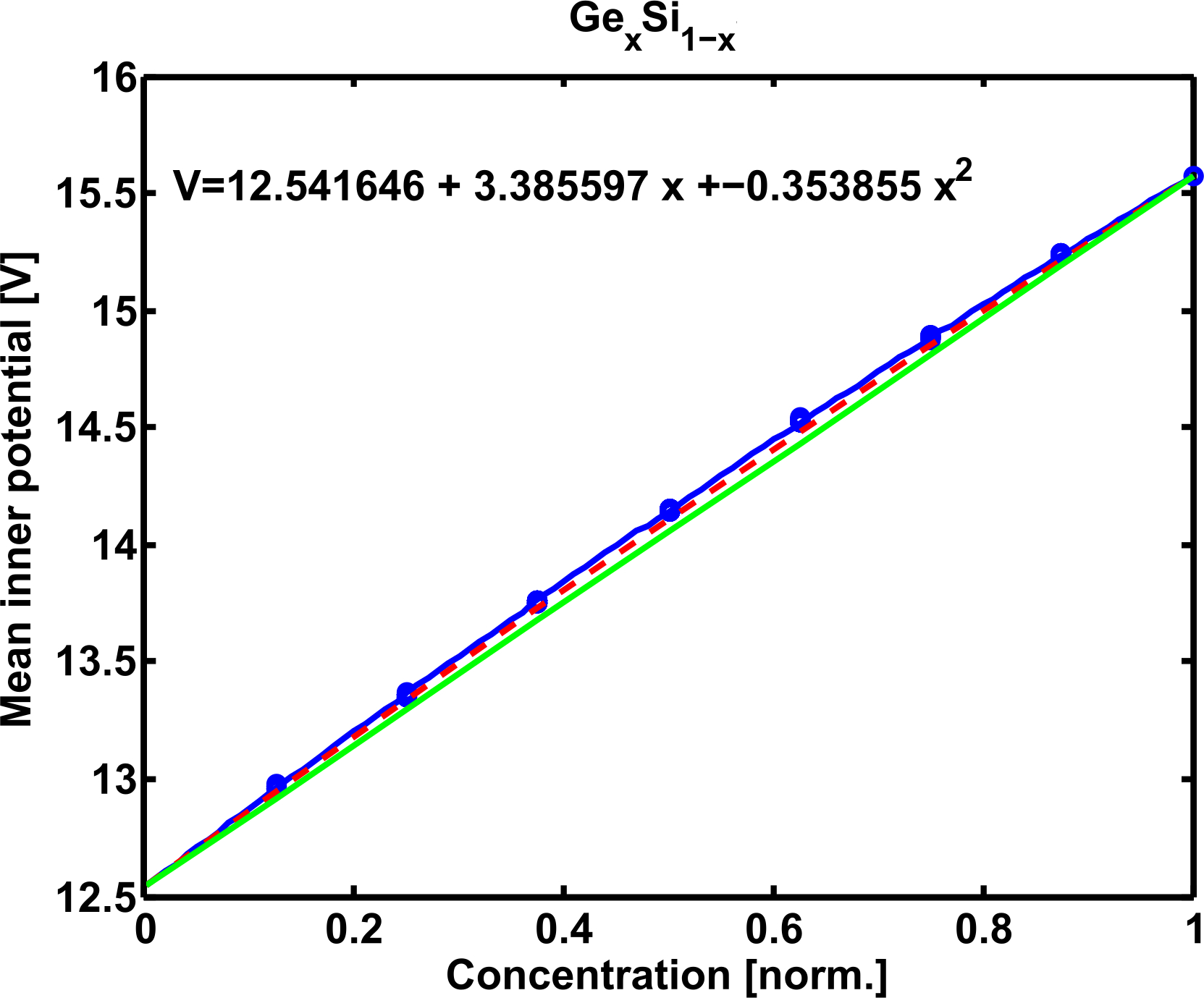}
    \caption{Computed MIP for Ge$_x$Si$_{1-x}$ biaxially stressed to Si in x and y direction and corresponding relaxation in z direction  plotted as blue circles. Solid blue lines are second order polynomial fits to the Boltzmann averaged MIPs. Green and red solid lines indicate the dependence assuming only linear interpolation and additionally volume rescaling.}
    \label{fig:MIP_alloy_biaxial}
\end{figure}

Remarkably, there is no significant difference between the more ionically bound Al$_x$Ga$_{1-x}$As and In$_x$Ga$_{1-x}$P systems compared to the more covalently bound Ge$_x$Si$_{1-x}$ system. Naivly, one would have expected a larger effect of the redistribution of charge due to alloying on the MIP of covalently compared to ionically bound materials. We attribute this to the significantly lower contribution of the interstitial regions to the MIP compared to the muffin-tin sphere regions (about 6 \% for GaP).

Finally, we tested the proposed model against a computation of Ge$_x$Si$_{1-x}$ biaxially stressed to Si in $x$ and $y$ direction and relaxed in $z$ direction (see fig. \ref{fig:MIP_alloy_biaxial}). The relaxation of the cell in $z$ was computed using elasticity theory employing linearly interpolated elastic moduli as given in tab.~\ref{tab:LattPara} for the unmixed materials. Again not a perfect agreement of our model with the computation was found. But it definitely better describes the dependence of the MIP on composition also for the cell biaxially stressed to Si as the pure linear interpolation between the MIPs of the pure materials. 

\begin{table}[h!]
    \centering
    \begin{tabular}{|c|c|c|c|c|}
         \hline
         Material& Lattice parameter [nm]& C$_{11}$ & C$_{12}$ & C$_{44}$\\
         \hline
         AlAs& 0.5661  & - & -& -  \\
         GaP & 0.5449  & - & -& -\\
         InP & 0.5868  & - & -& -\\
         Si & 0.54311  & 163.8 & 59.2& 81.7\\
         Ge & 0.565791 & 126.0 & 44.0 & 67.7\\
         \hline
    \end{tabular}
    \caption{Lattice parameter and elastic moduli (in GPa) as used for the computations.}
    \label{tab:LattPara}
\end{table}

Finally, we want to state that the model also predicts the mass density dependence of the MIP of amorphous carbon. In reference \cite{SchowalterAmorphC} the MIP of amorphous carbon was computed for various mass densities. An approximately linear relation was found and fitted by 
$$
V_0(\rho)=5.20\frac{\text{V}}{\text{g/cm}^3} \rho+0.76 \text{V}\quad\text{.}
$$
Using the proposed model the MIP would be proportional to the mass density with vanishing MIP at vanishing mass density. The slope would be given for a density $\rho_\text{ref}=3.5 \text{g/cm}^3$ by $m=\frac{V_0(\rho_\text{ref})}{\rho_\text{ref}}=\frac{19 \text{V}}{3.5\text{g/cm}^3}=5.43 \frac{\text{V}}{\text{g/cm}^3}$. Our model predicts a larger slope than it was found from the fit of the computed values several years ago. However, the fit allowed for a non-vanishing y-intercept, which our model does not. Fixing the y-intercept at the origin results in a larger slope of 5.47 $\frac{\text{V}}{\text{g/cm}^3}$, which is very close to our model's value. Thus, we conclude that our model also describes the mass density dependence comparably well.  

In summary, we showed that dependence of the MIP of mixed materials on concentration rather well can be described by linear interpolation as long as the volume stays unchanged with concentration. This is due to the fact, that the largest contribution to the MIP arises from the large Coulomb potential within the muffin-tin spheres and thus, becomes linear by definition. For dependencies with changing volumes (lattice parameter differences of pure materials, strain, density changes, etc.) the MIP needs to be rescaled by the volume, as the integral over the Coulomb potential (eq. (\ref{eq:MIP_def})) stays approximately constant and thus, the MIP is mainly changed by the change in volume. In any case, the proposed scheme better describes the MIP than a simple linear interpolation. It can be as easily applied to described situations, as a sole linear interpolation. 

%


\begin{acknowledgments}
M.S. acknowledges computation time on the LESUM cluster of the University Bremen.

\end{acknowledgments}

\section*{Data Availability Statement}

The data that support the findings of this study are deposited on Zenodo \cite{Zenodo}.

\bibliography{MIP}

\end{document}